\documentclass[aps,reprint,superscriptaddress,preprintnumbers,longbibliography,nofootinbib]{revtex4-1}

\usepackage{amsthm}
\usepackage{graphicx}
\usepackage{amsmath}
\usepackage{hyperref}
\usepackage{xspace}
\usepackage{braket}
\usepackage{todonotes}
\usepackage{subcaption}

\usepackage{dcolumn}
\usepackage{bm}

\newtheorem*{theorem*}{Theorem}

\DeclareRobustCommand{\Sec}[1]{Sec.~\ref{#1}}

\DeclareRobustCommand{\Ref}[1]{Ref.\,\cite{#1}}
\DeclareRobustCommand{\Refs}[1]{Refs.\,\cite{#1}}

\usepackage{color}
\definecolor{darkblue}{rgb}{0,0,0.5}
\definecolor{darkred}{rgb}{0.5,0,0}
\definecolor{darkgreen}{rgb}{0,0.5,0}

\newcommand{\be}{\begin{equation}}
\newcommand{\ee}{\end{equation}}

\allowdisplaybreaks

\begin{document}

\preprint{SLAC-PUB-17644}

\title{A General Analysis for Observing Quantum Interference at Colliders}

\author{Andrew J.~Larkoski}
\email{larkoski@slac.stanford.edu}
\affiliation{SLAC National Accelerator Laboratory, Menlo Park, CA 94025, USA}

\date{\today}

\begin{abstract}
\noindent Despite their inextricable quantum mechanical nature, events at a high energy particle collider experiment typically have very few unambiguous quantum signatures, due the type of data and the manner in which they are collected.  We present a general analysis of one feature of quantum mechanics, interference between two orthogonal states on Hilbert space, projected onto the basis of states that span a collider experiment observable space.  Identification of quantum interference can be considered as a binary discrimination between a pure state and a mixed state, and we introduce several statistical measures that quantify the amplitude of interference in a pure state with respect to a mixed state that exhibits no interference.  Two explicit examples from particle physics are provided to demonstrate features of the general formalism. 
\end{abstract}

\maketitle

\tableofcontents

\section{Introduction}

Though the physics at particle collision experiments is governed by quantum mechanics, the type of measurements and the way in which they are performed often significantly obscures observable quantum effects.  Entanglement is perhaps the most shocking prediction of quantum mechanics to a classical physicist, but only in limited situations at a collider can entanglement potentially be observed.  In general, the quantum physics at a collider like the Large Hadron Collider is at a distance scale of a femtometer, and detectors are centimeters or meters from the point of collision.  So, a particle detector essentially lives on the celestial sphere, both an infinite distance and time from the point at which particles were produced.  As such, distinct cells in a calorimeter or pixels in a tracker are space-like separated, and exhibit no causal connection.  Therefore, measurements of particles at different locations in the detector commute.

Conclusively demonstrating quantum entanglement beyond simply classical correlations can be accomplished through inequalities, like Bell's inequalities \cite{Bell:1964kc}.  However, for a quantum system to violate Bell's inequalities, distinct measurements must not commute.  While this isn't necessarily a no-go theorem, the way in which measurements are performed at a collider present a significant barrier to demonstrating entanglement in general collision processes.  Nevertheless, techniques have been developed to test Bell's equalities in astrophysical data \cite{Handsteiner:2016ulx}, or in thought experiments for cosmological tests \cite{Maldacena:2015bha}.   Additionally, there have been recent efforts at colliders to observe entanglement in exploiting the left-handed of top quark pair production and decay, in spin correlations from Higgs boson decay, and in flavor and spin correlations in hadron decays \cite{Afik:2020onf,Fabbrichesi:2021npl,Takubo:2021sdk,Severi:2021cnj,Barr:2021zcp,Gong:2021bcp}.

Another signature of quantum mechanics is interference.  Quantum interference arises from coherently summing over multiple orthogonal states on Hilbert space that are all consistent with the performed measurement.  With a general basis on Hilbert space, quantum interference is manifest as off-diagonal elements in the density matrix.  When the density matrix is subsequently projected onto the basis of states observed in experiment, the off-diagonal elements are manifest as constructive or destructive interference in a probability distribution.  The presence of this source of interference cannot be reproduced classically, but is always present in any experiment when multiple orthogonal states are consistent with measurement.  Observing quantum interference in a particle physics experiment is therefore much more feasible than observing entanglement.  In this paper, we will present the formalism for quantum interference relevant for collider physics, and present techniques for its observation.

It is important to emphasize both what quantum interference is and what it is not.  Quantum interference only exists if your measurement is inclusive over two or more orthogonal states on Hilbert space.  For example, in the double-slit experiment, if your measurement exclusively consists of viewing the screen, you are completely ignorant of which slit photons pass through.  The state of a photon passing through one slit is orthogonal to the state where the photon passes through the other slit.  Hence, you observe an interference pattern on the screen.  However, if you also place a detector at one of the slits, then the only way that your measurement can be consistent is if a photon passes through that slit.  Hence, there is but a single quantum state consistent with your measurements and so no interference pattern is observed on the screen.  To potentially observe quantum interference in a collider therefore requires identical detected final states, but produced from multiple, orthogonal paths from the initial state.

Because Feynman diagrams are typically the way that predictions for collision events are calculated, it is worth considering the distinction between summing over Feynman diagrams and quantum interference.  An intermediate state is necessarily physical; it must live on Hilbert space.  In a gauge theory, this means that a state must be gauge invariant, for example.  In calculations with Feynman diagrams, typically multiple diagrams with distinct topology, but identical external particles, must be included.  Such a sum then produces the amplitude for the desired scattering process to occur, which can be squared to determine the distribution on phase space, for example.  However, the coherent sum over Feynman diagrams does not mean that the diagrams ``interfere'' in the quantum sense we take here.  Again, in a gauge theory, only the sum of diagrams is gauge invariant and therefore physical.  Depending on your choice of gauge, the representation of intermediate gauge bosons is gauge dependent and therefore the value of any individual Feynman diagram is also gauge dependent.  We will present an example of quantum interference of distinct spin states of an intermediate gluon, but to do so we must force a kinematic regime of the final state particles where the gluon becomes real.

Efforts to identify quantum interference at a high-energy collider are rather limited in the literature.  Quantum interference of orthogonal spin states of a new particle resonance was identified as a viable technique for establishing the new particle's spin \cite{Buckley:2007th,Buckley:2008pp} (see also \Ref{Gottfried:1964nx}).  Subsequent analyses typically just focused on applications to identification of the spin of a hypothetical particle in some new model of physics beyond the Standard Model (e.g., \Refs{Alves:2008up,Boudjema:2009fz,Murayama:2009jz}).  Recently, an observable for detecting quantum interference between the helicity states of a gluon was introduced \cite{Chen:2020adz}, which we review in \Sec{sec:ex}.  Further, a parton shower algorithm has been developed and validated that correctly includes spin interference effects \cite{Karlberg:2021kwr,Hamilton:2021dyz}, and such interference effects were recognized long ago as vital for understanding collider processes \cite{Webber:1986mc,Collins:1987cp,Knowles:1987cu}.  However, to the best of our knowledge, a general discussion and analysis for quantum interference at a collider has not been done, and is the central goal of this paper.

This paper is organized as follows.  In \Sec{sec:pure}, we present a formalism for quantum interference in a particle physics experiment that only measures particle momenta, and no other quantum numbers.  We define a pure state from two interfering states as well as the corresponding mixed state, and construct the optimal observable for their discrimination.  In \Sec{sec:tech}, we define and provide robust bounds on several statistical measures for establishing quantum interference from the pure and mixed state distributions of the discrimination observable.  Through this section, all results will be completely general, given the established assumptions for measurements at a collider.  In \Sec{sec:ex}, we present physical examples of interference that illustrate the general features and for which could potentially be searched in collider data.    We conclude and look forward in \Sec{sec:conc}.

\section{Pure and Mixed States on Phase Space}\label{sec:pure}

Consider a pure state $|\psi\rangle$ formed from the linear combination of two orthogonal states $|A\rangle$ and $|B\rangle$ on Hilbert space:
\begin{align}
|\psi\rangle = \frac{1}{\sqrt{1+\epsilon}}|A\rangle+\sqrt{\frac{\epsilon}{1+\epsilon}}\,|B\rangle\,.
\end{align}
These states are normalized: $\langle A|A\rangle = \langle B|B\rangle =1$.  We assume that $\epsilon\in[0,1]$ is their mixing and orthogonality is $\langle A|B\rangle = 0$.  At a collider experiment, each event consists of a fixed number of particles and we measure their momenta, so we observe these states projected onto the Fock space of $N$-particle momentum states.  An $N$-particle momentum state $|k_1,\dotsc,k_N\rangle$ satisfies the completeness relation
\begin{align}
1 = \sum_{N=0}^\infty \frac{1}{N!}\int d\Pi_N\, |k_1,\dotsc,k_N\rangle\langle k_1,\dotsc,k_N|\,,
\end{align}
where $d\Pi_N$ is differential $N$-body Lorentz-invariant phase space and the $1/N!$ normalization arises from assuming the particles are indistinguishable.  That is, we will also assume that we only measure the particle momenta; no other quantum numbers are observed.  In this space, the states are just scattering amplitudes:
\begin{align}
{\cal A}_N(A)&\equiv \langle k_1,\dotsc,k_N|A\rangle\,,\\
{\cal A}_N(B)&\equiv \langle k_1,\dotsc,k_N|B\rangle\,.
\end{align}
Orthogonality in $N$-particle momentum space is manifest through integration over phase space:
\begin{align}
0 = \langle A|B\rangle =  \int d\Pi_N\, {\cal A}_N^*(A){\cal A}_N(B)\,.
\end{align}
When conditioned on the number of observed particles $N$, the squared amplitudes are also normalized.  In examples we consider later, the orthogonal states on Hilbert space that interfere will only have non-zero projection onto a single $N$-particle state.  We will therefore assume in the following that the squared amplitudes are normalized for fixed $N$ particle number.  

At each phase space point defined by a unique state $|k_1,\dotsc,k_N\rangle$, we can define a $2\times 2$ density matrix $\rho_\text{pure}$ as
\begin{align}
\rho_\text{pure} &=|\psi\rangle\langle \psi|= \frac{1}{1+\epsilon}|A\rangle\langle A| + \frac{\epsilon}{1+\epsilon}|B\rangle\langle B| \\
&
\hspace{3cm}+ \frac{\sqrt{\epsilon}}{1+\epsilon}\left(
|A\rangle\langle B| + |B\rangle\langle A|
\right)\,.\nonumber
\end{align}
The probability density on $N$-particle momentum space is therefore
\begin{align}
&d\Pi_N\,\langle k_1,\dotsc,k_N|\rho_\text{pure}|k_1,\dotsc,k_N\rangle \\
&
\hspace{1cm}
= d\Pi_N\, \left[
\frac{1}{1+\epsilon}|{\cal A}_N(A)|^2+\frac{\epsilon}{1+\epsilon}|{\cal A}_N(B)|^2 \right.\nonumber\\
&
\hspace{3.5cm}\left.+ 2\frac{\sqrt{\epsilon}}{1+\epsilon}\,\text{Re}\left(
{\cal A}_N^*(A){\cal A}_N(B)
\right)
\right]\,.\nonumber
\end{align}

We can also define a mixed state that exhibits no quantum interference at each phase space point, which has density matrix $\rho_\text{mix}$:
\begin{align}
\rho_\text{mix} = \frac{1}{1+\epsilon}|A\rangle\langle A| + \frac{\epsilon}{1+\epsilon}|B\rangle\langle B|\,.
\end{align}
Its probability density on phase space is then
\begin{align}
&d\Pi_N\,\langle k_1,\dotsc,k_N|\rho_\text{mix}|k_1,\dotsc,k_N\rangle \\
&\hspace{1cm}= d\Pi_N\, \left[
\frac{1}{1+\epsilon}|{\cal A}_N(A)|^2+\frac{\epsilon}{1+\epsilon}|{\cal A}_N(B)|^2 
\right]\,.\nonumber
\end{align}
Our goal in this paper will be to establish general properties of these two distributions and attempt to maximally distinguish them, with the ultimate goal of unambiguously observing quantum interference.

\subsection{Maximally-Discriminating Observable}

By the Neyman-Pearson Lemma \cite{Neyman:1933wgr}, the optimal observable for discrimination of these pure and mixed states on phase space is their likelihood ratio ${\cal L}$:
\begin{align}
{\cal L} &\equiv \frac{\langle k_1,\dotsc,k_N|\rho_\text{pure}|k_1,\dotsc,k_N\rangle}{\langle k_1,\dotsc,k_N|\rho_\text{mix}|k_1,\dotsc,k_N\rangle} \\
&=1+\frac{2\sqrt{\epsilon}\,\text{Re}\left(
{\cal A}_N^*(A){\cal A}_N(B)
\right)}{|{\cal A}_N(A)|^2+\epsilon|{\cal A}_N(B)|^2 }\,.\nonumber
\end{align}
We can also express the likelihood as
\begin{align}
{\cal L}  =1+\frac{2\sqrt{\epsilon}\,
|{\cal A}_N(A)|\,|{\cal A}_N(B)|}{|{\cal A}_N(A)|^2+\epsilon|{\cal A}_N(B)|^2 }\cos\phi\,,
\end{align}
where $\phi$ is the argument of ${\cal A}_N^*(A){\cal A}_N(B)$.  This form then makes it clear that the maximal possible range of ${\cal L}$ is
\begin{align}
{\cal L}\in [0,2]\,.
\end{align}
At ${\cal L} = 0$, only the mixed state can exist, while at ${\cal L} = 2$, the pure state is twice as likely as the mixed state.  The actual physical range of ${\cal L}$ will depend in detail on the particular form of the interfering amplitudes ${\cal A}_N(A)$ and ${\cal A}_N(B)$ and how they vary over phase space.  For convenience later, we will shift the likelihood ratio by 1, so that it ranges from $[-1,1]$, which has no effect on its discrimination power.  We call
\begin{align}
{\cal O} &\equiv  \frac{\langle k_1,\dotsc,k_N|\rho_\text{pure}-\rho_\text{mix}|k_1,\dotsc,k_N\rangle}{\langle k_1,\dotsc,k_N|\rho_\text{mix}|k_1,\dotsc,k_N\rangle}\\
&=\frac{2\sqrt{\epsilon}\,\text{Re}\left(
{\cal A}_N^*(A){\cal A}_N(B)
\right)}{|{\cal A}_N(A)|^2+\epsilon|{\cal A}_N(B)|^2 }\in[-1,1]\,.\nonumber
\end{align}
This form manifests sensitivity to off-diagonal elements of the pure state density matrix via
\begin{align}
\rho_\text{pure}-\rho_\text{mix} = \frac{\sqrt{\epsilon}}{1+\epsilon}\left(
|A\rangle\langle B| + |B\rangle\langle A|
\right)\,.
\end{align}

With this observable, we can then calculate its probability distribution on $N$-body phase space, for the pure and mixed states.  From the probability densities established above, we have
\begin{align}
p_\text{pure}({\cal O}) &= \int d\Pi_N\,  \left(
\frac{1}{1+\epsilon}|{\cal A}_N(A)|^2+\frac{\epsilon}{1+\epsilon}|{\cal A}_N(B)|^2 \right.\nonumber\\
&\left.
\hspace{1.5cm}+\frac{2\sqrt{\epsilon}}{1+\epsilon}\,\text{Re}\left(
{\cal A}_N^*(A){\cal A}_N(B)
\right)
\right)\\
&
\hspace{0.25cm}
\times\delta\left(
{\cal O} -\frac{2\sqrt{\epsilon}\,\text{Re}\left(
{\cal A}_N^*(A){\cal A}_N(B)
\right)}{|{\cal A}_N(A)|^2+\epsilon|{\cal A}_N(B)|^2 }
\right)\,,\nonumber\\
p_\text{mix}({\cal O}) &= \int d\Pi_N\,  \left(
\frac{1}{1+\epsilon}|{\cal A}_N(A)|^2+\frac{\epsilon}{1+\epsilon}|{\cal A}_N(B)|^2 
\right)\nonumber\\
&
\hspace{0.25cm}\times\delta\left(
{\cal O} -\frac{2\sqrt{\epsilon}\,\text{Re}\left(
{\cal A}_N^*(A){\cal A}_N(B)
\right)}{|{\cal A}_N(A)|^2+\epsilon|{\cal A}_N(B)|^2 }
\right)\,.\label{eq:mixdist}
\end{align}
By the orthogonality of states $|A\rangle$ and $|B\rangle$, the expectation value of ${\cal O}$ on the mixed state is 0:
\begin{align}
\int_{-1}^1 d{\cal O}\, {\cal O}\, p_\text{mix}({\cal O}) = 0\,.
\end{align}
Additionally, note that the pure state distribution is related to the mixed state distribution as:
\begin{align}
p_\text{pure}({\cal O}) = (1+{\cal O})\, p_\text{mix}({\cal O})\,,
\end{align}
using the $\delta$-function to exchange the explicit mixing term in the probability density for the observable ${\cal O}$.  Using this relationship, we can establish general bounds for discrimination and hypothesis testing.

In general, not much more can be said about the properties of the pure and mixed state distributions of ${\cal O}$.  A property of the mixed state distribution that might be suggested by its form on phase space is symmetry about ${\cal O} = 0$, for which $p_\text{mix}(-{\cal O})=p_\text{mix}({\cal O})$.  This is indeed true for some interesting cases that we will review in examples later, but is not true in general.  We will construct an example for which the mixed state distribution is not symmetric, with only the assumptions made so far.  To ensure that the mixed state distribution is symmetric, more restrictions on the relationship between the states $|A\rangle$ and $|B\rangle$ must be enforced.

%

\section{Techniques for Identifying Quantum Interference}\label{sec:tech}

With the established pure and mixed state probability distributions for the interference observable ${\cal O}$, we are in a position to define techniques for establishing quantum interference and quantify the discrimination power between these two systems.  In this section, we will present a number of statistical measures, their physical interpretation, and prove general bounds on their values given the properties of the mixed and pure state distributions.  In the following section, we will present explicit collider physics examples of quantum interference and calculate these quantities.

\subsection{Destructive Interference Limit}

Unlike the mixed state, the pure state can potentially exhibit complete destructive interference, when the mixing between states $|A\rangle$ and $|B\rangle$ is maximal, $\epsilon = 1$, and their phases differ by $\phi = \pi$.  At this destructive interference point, ${\cal O} = -1$, and necessarily $p_\text{pure}({\cal O} = -1) = 0$, because the probability density on phase space vanishes at that point.  By contrast, the mixed state distribution is not required to vanish at ${\cal O} = -1$, as
\begin{align}
p_\text{mix}({\cal O}) = \frac{p_\text{pure}({\cal O} )}{1+{\cal O}}\,.
\end{align}
By integrability of the probability distributions, all that is required as ${\cal O}\to -1$ is that the pure state distribution vanishes.  However, this characteristic of the pure state distribution is subtle and may not be directly useful if the states $|A\rangle$ and $|B\rangle$ do not exhibit maximal mixing: $\epsilon < 1$.  In this situation, both the pure and mixed state distributions can vanish at and below some ${\cal O}_{\min} >-1$.  We will encounter a physical example of this case in the next section.

\subsection{Purity on $N$-Body Phase Space}

Given a density matrix of some quantum system $\rho$, there are several established quantities that can be used as a measurement of the amount of mixing.  Perhaps the most familiar is the von Neumann entropy $S_\text{vN}$ for which
\begin{align}
S_\text{vN} = -\text{tr}\left(
\rho\log \rho
\right)\,.
\end{align}
The von Neumann entropy satisfies (strong) subadditivity and other powerful relations \cite{kiefer1959optimum,Araki:1970ba,Lieb:1973cp} that no other entropy-like measure can because of unique properties of the logarithm.  However, the von Neumann entropy can be challenging to calculate in practice because evaluating $\log \rho$ requires diagonalizing the density matrix, which may be very inconvenient if the density matrix is not diagonal in the natural space of your measurements.

Nevertheless, there are simple relationships of the density matrix $\rho$ which exploit properties of a pure state, in contrast to a mixed state.  For example, exclusively for a pure state is the density matrix idempotent:
\begin{align}
\rho_\text{pure}^2 = \rho_\text{pure}\,.
\end{align}
By conservation of probability, $\text{tr}\,\rho = 1$ and $\rho$ is a Hermitian matrix and so the linear entropy $S_L$ measures the level of mixture in the system:
\begin{align}
S_L = \text{tr}(\rho-\rho^2) = 1-\text{tr}\,\rho^2\,.
\end{align}
When projected onto $N$-particle momentum states, the linear entropy is
\begin{align}
S_L = \sum_{N=0}^\infty\frac{1}{N!} \int d\Pi_N\, \langle k_1,\dotsc,k_N|\rho-\rho^2|k_1,\dotsc,k_N\rangle\,.
\end{align}
Assuming completeness of $N$-particle momentum states, this vanishes on a pure state, $S_L = 0$.

For our example of a mixed state defined by orthogonal states $|A\rangle$ and $|B\rangle$, note that the square of the density matrix is
\begin{align}
\rho_\text{mix}^2 = \frac{1}{(1+\epsilon)^2}|A\rangle\langle A| + \frac{\epsilon^2}{(1+\epsilon)^2}|B\rangle\langle B|\,,
\end{align}
and so the difference of the density matrix and its square is
\begin{align}
\rho_\text{mix}-\rho_\text{mix}^2=\frac{\epsilon}{(1+\epsilon)^2}\left(|A\rangle\langle A| + |B\rangle\langle B|\right)\,.
\end{align}
The linear entropy of this system is then a direct measurement of the mixing $\epsilon$, with
\begin{align}
S_L &= \text{tr}(\rho_\text{mix}-\rho_\text{mix}^2)= \frac{\epsilon}{(1+\epsilon)^2}\left(
\langle A|A\rangle + \langle B|B\rangle
\right)\\
&=\frac{2\epsilon}{(1+\epsilon)^2}\,.\nonumber
\end{align}

While the linear entropy is interesting as a measure of mixing of the states $|A\rangle$ and $|B\rangle$, it is of limited use on measurements at a collider experiment.  Any entropy measure requires knowledge of the density matrix of the system, and direct measurement of the off-diagonal elements of the density matrix is generally not possible.  All quantities measured at a collider are probability distributions on $N$-body phase space, and are necessarily diagonal in the basis of $N$-body momentum states.  There have been some recent developments in defining idealized observables that can probe off-diagonal elements of the density matrix in the $N$-particle momentum basis; for example, \Ref{Korchemsky:2021okt}.  These examples, however, are not practically realizable, as they require continuous knowledge of the particles, from the point of scattering to infinity, to ensure that subsequent measurements can be causally connected and therefore do not in general commute.

\subsection{Area Under the ROC Curve}

For any binary discrimination problem, a useful way to express the efficiency of the discriminating observable is through the receiver operating characteristic (ROC) curve, which displays the true positive rate as a function of the false positive rate.  A single number that quantifies the efficacy of the discriminant from the ROC curve is the area under it (AUC).  For perfect discrimination, the true positive rate is 1 for all values of the false positive rate; therefore, perfect discrimination has $\text{AUC} = 1$.  By contrast, a completely random discriminant assigns true positives and false positives to events at an equal rate, and so the AUC for the worst possible discriminant is $\text{AUC} = 1/2$.  We will establish bounds on the AUC for discrimination of pure versus mixed states on phase space.

For the interference observable ${\cal O}$ on the pure and mixed distributions, the AUC is defined as the ordered integral:
\begin{align}
\text{AUC} &=   \int_{-1}^1 d{\cal O}_1 \int_{{\cal O}_1}^1 d{\cal O}_2\, p_\text{mix}({\cal O}_1)\,p_\text{pure}({\cal O}_2)\\
& =\int_{-1}^1 d{\cal O}_1 \int_{{\cal O}_1}^1 d{\cal O}_2\, (1+{\cal O}_2)\,p_\text{mix}({\cal O}_1)\,p_\text{mix}({\cal O}_2)\nonumber\\
&=\int_{-1}^1 d{\cal O}\, (1+{\cal O})\,\Sigma_\text{mix}({\cal O})\,p_\text{mix}({\cal O})\nonumber\,,
\end{align}
where $\Sigma_\text{mix}({\cal O})$ is the cumulative distribution function, and we have used the functional relationships between the pure and mixed distributions of ${\cal O}$.  Note that
\begin{align}
\int_{-1}^1 d{\cal O}\, \Sigma_\text{mix}({\cal O})\,p_\text{mix}({\cal O}) = \frac{1}{2}\,,
\end{align}
using integration by parts and normalization of the probability distribution.  Also, note that
\begin{align}
\int_{-1}^1 d{\cal O}\, {\cal O}\,\Sigma_\text{mix}({\cal O})\,p_\text{mix}({\cal O}) =\frac{1}{2} - \frac{1}{2}\int_{-1}^1d{\cal O}\, \Sigma^2_\text{mix}({\cal O})\,,
\end{align}
again using integration by parts.  Then, the AUC is
\begin{align}
\text{AUC} = 1-\frac{1}{2}\int_{-1}^1d{\cal O}\, \Sigma^2_\text{mix}({\cal O})\,.
\end{align}

Because the expectation value of ${\cal O}$ is zero on the mixed state distribution, the cumulative distribution integrates to 1:
\begin{align}
\int_{-1}^1d{\cal O}\, \Sigma_\text{mix}({\cal O}) = 1\,,
\end{align}
using integration by parts.  Also, because the cumulative distribution is non-negative and bounded from above by 1, 
\begin{align}
\Sigma_\text{mix}({\cal O})\geq \Sigma^2_\text{mix}({\cal O})\,.
\end{align}
It then follows that 
\begin{align}
\text{AUC} &= 1-\frac{1}{2}\int_{-1}^1d{\cal O}\, \Sigma^2_\text{mix}({\cal O})\\
&\geq 1-\frac{1}{2}\int_{-1}^1d{\cal O}\, \Sigma_\text{mix}({\cal O}) = \frac{1}{2}\,,\nonumber
\end{align}
which is simply the statement that the AUC must at least be the value corresponding to a completely ineffective discriminant.

A non-trivial upper bound on the AUC can be established using the properties of the ROC curve for the likelihood ratio.  For the likelihood ratio (or any monotonic function of it), the ROC curve is both monotonic and concave-down, and therefore can be bounded by the area of a quadrilateral \cite{Larkoski:2019nwj}.  The area of this quadrilateral is related to the maximum and minimum values of the likelihood ratio:
\begin{align}
\text{AUC}\leq \frac{2{\cal L}_{\max}-1+{\cal L}_{\max}{\cal L}_{\min}}{2({\cal L}_{\max}-{\cal L}_{\min})}\,.
\end{align}
We had established that the maximum and minimum values of the likelihood ratio are
\begin{align}
&{\cal L}_{\max} = 2\,, &{\cal L}_{\min} = 0\,,
\end{align}
and so the AUC is bounded as
\begin{align}
\frac{1}{2}\leq \text{AUC}\leq \frac{3}{4}\,.
\end{align}
Interestingly, this also places a lower bound on the integral of the square of the cumulative distribution of the mixed state:
\begin{align}
\frac{1}{2}\leq \int_{-1}^1d{\cal O}\, \Sigma^2_{\text{mix}}({\cal O}) \leq 1\,.
\end{align}

The mixed state distribution with the most perfect discrimination that saturates these bounds corresponds to:
\begin{align}\label{eq:maxdist}
p_\text{mix}^{\max}({\cal O}) = \frac{1}{2}\left(
\delta(1+{\cal O})+\delta(1-{\cal O})
\right)\,.
\end{align}
The corresponding pure state distribution is then
\begin{align}
p_\text{pure}^{\max}({\cal O}) = (1+{\cal O})p^{\max}_\text{mix}({\cal O}) = \delta(1-{\cal O})\,.
\end{align}
For this distribution, note that its cumulative distribution is constant on ${\cal O}\in(-1,1)$:
\begin{align}
\Sigma_\text{mix}^{\max}({\cal O}) = \frac{1}{2}\,,
\end{align}
and therefore its AUC would be
\begin{align}
\text{AUC}^{\max} = 1-\frac{1}{2}\int_{-1}^1 d{\cal O}\, \Sigma^{\max}_\text{mix}({\cal O})^2 = \frac{3}{4}\,,
\end{align}
satisfying the general bound.

\subsection{Kolmogorov-Smirnov test}

For hypothesis testing, the Kolmogorov-Smirnov (KS) test is a useful metric and its evaluation is straightforward for these observables.  The KS test is the maximum difference between the cumulative distributions of the null hypothesis and data.  In this quantum interference example, we take the null hypothesis to be the mixed state distribution and the ``data'' to be the pure state distribution.  So, we have
\begin{align}
\text{KS} &= \text{sup}_{{\cal O}} \left|
\Sigma_\text{mix}({\cal O})-\Sigma_\text{pure}({\cal O})
\right| \\
&= \text{sup}_{{\cal O}} \left|\int_{-1}^{{\cal O}} d{\cal O}{'}\, 
\left(1-(1+{\cal O}{'})\right)
p_\text{mix}({\cal O}{'})\right|\nonumber\\
&=\text{sup}_{{\cal O}} \left|\int_{-1}^{{\cal O}} d{\cal O}{'}\, 
{\cal O}{'}
p_\text{mix}({\cal O}{'})\right|\,.\nonumber
\end{align}
This is maximized if ${\cal O}=0$.  Then,
\begin{align}
\text{KS} &= -\int_{-1}^0 d{\cal O}\, 
{\cal O}\,
p_\text{mix}({\cal O}) = \int_{-1}^0 d{\cal O}\, 
\Sigma_\text{mix}({\cal O})\,.
\end{align}
Because the cumulative distribution is monotonic and its integral is 1 over the entire domain of ${\cal O}$, there is a non-trivial upper bound to the KS statistic.  At most, half of the integral of the cumulative distribution can lie in the domain ${\cal O}\in[-1,0]$ and so
\begin{equation}
0\leq \text{KS}\leq \frac{1}{2}\,.
\end{equation}

The value of the KS test also has an interesting interpretation as a measure of the net interference in the pure state.  The term in the pure state distribution $p_\text{pure}({\cal O})$ that arises exclusively from the interference is
\begin{align}
p_{\text{pure}}({\cal O})-p_{\text{mix}}({\cal O}) = {\cal O}\,p_{\text{mix}}({\cal O})\,.
\end{align}
Of course, when integrated on all of ${\cal O}\in[-1,1]$, this vanishes, but restricting to ${\cal O}\in[0,1]$ ensures that there are only non-negative contributions to the interference.  So, a measure of the amplitude of the interference is:
\begin{align}
\text{A}_\text{FB,pure} &\equiv \int_0^1 d{\cal O}\, {\cal O}\,p_\text{mix}({\cal O})\\
&=1-\int_0^1d{\cal O}\,\Sigma_\text{mix}({\cal O})\,.\nonumber
\end{align}
We refer to this as a ``forward-backward asymmetry'' because it is half of the difference of the integral of ${\cal O}\,p_{\text{mix}}({\cal O})$ from ${\cal O}\in[0,1]$ to ${\cal O}\in[-1,0]$.  Because the integral of the cumulative distribution on ${\cal O}\in[-1,1]$ is 1, this is also
\begin{align}
\text{A}_\text{FB,pure} &=\int_{-1}^0d{\cal O}\,\Sigma_\text{mix}({\cal O}) = \text{KS}\,.
\end{align}

\subsection{Kullback-Leibler Divergence}

The Kullback-Leibler (KL) divergence is a quantity that measures the amount of information required to describe one distribution with respect to another.  The KL divergence of the pure state distribution with respect to the mixed state distribution is
\begin{align}
D_\text{KL}(p_\text{pure}||p_\text{mix}) &= \int_{-1}^1 d{\cal O}\, p_\text{pure}({\cal O}) \log\frac{p_\text{pure}({\cal O})}{p_\text{mix}({\cal O})}\\
&=\int_{-1}^1 d{\cal O}\, p_\text{mix}({\cal O})\, (1+{\cal O}) \log(1+{\cal O})\,.\nonumber
\end{align}
Note that the KL divergence is only defined for the pure distribution with respect to the mixed distribution so that the limit when a distribution vanishes makes sense.

The KL divergence is non-negative, and takes 0 value if the pure and mixed state distributions are identical.  That limit corresponds to the distributions both $\delta$-functions peaking at ${\cal O} = 0$:
\begin{align}
p_\text{pure}({\cal O})=p_\text{mix}({\cal O}) = \delta({\cal O})\,.
\end{align}
The maximum value of the KL divergence can be established using the optimal discrimination distribution from Eq.~\ref{eq:maxdist}.  On this distribution, the KL divergence takes the value
\begin{align}
D^{\max}_\text{KL}(p_\text{pure}||p_\text{mix}) &= \int_{-1}^1 d{\cal O}\, p^{\max}_\text{mix}({\cal O})(1+{\cal O}) \log(1+{\cal O})\nonumber\\
&=\log 2\,.
\end{align}
That is, the maximum amount of information encoded exclusively in the interference of states $|A\rangle$ and $|B\rangle$ is $\log 2$.  In general, the KL divergence is less than this:
\begin{align}
0 \leq D_\text{KL}(p_\text{pure}||p_\text{mix}) \leq \log 2\,.
\end{align}

\section{Examples}\label{sec:ex}

We now turn to consideration of explicit examples to demonstrate the general results established above.  We will present three examples in this section: the first is just the form for possible interference, while the latter two are collider physics examples, that can in principle be searched for.

\subsection{Sinusoidal Interference}

Let's consider the simple example of two states whose magnitude on phase space are both constants, but have non-trivial interference.  That is, we will take the probability distributions to be
\begin{align}
&p_\text{mix}(\phi) = \frac{1}{\pi} \,,
&p_\text{pure}(\phi) = \frac{1}{\pi} \left(1+\epsilon \cos\phi\right)\,,
\end{align}
where $\phi\in[0,\pi]$ is their relative phase and $\epsilon\in[0,1]$.  The discrimination observable is
\begin{align}
{\cal O} = \epsilon \cos\phi\,,
\end{align}
and its distributions on the mixed and pure states are
\begin{align}
p_\text{mix}({\cal O}) &= \frac{1}{\pi}\frac{1}{\sqrt{\epsilon^2-{\cal O}^2}}\,,\\
p_\text{pure}({\cal O}) &= \frac{1}{\pi}\frac{1+{\cal O}}{\sqrt{\epsilon^2-{\cal O}^2}}\,.\nonumber
\end{align}
The cumulative distribution for the mixed distribution is
\begin{align}
\Sigma_\text{mix}({\cal O}) = \frac{1}{\pi}\tan^{-1}\left(
\frac{{\cal O}}{\sqrt{\epsilon^2-{\cal O}^2}}
\right)+\frac{1}{2}\,,
\end{align}
where $-\epsilon\leq {\cal O}\leq \epsilon$.

It then follows that the AUC is
\begin{align}
\text{AUC} = 1-\frac{1}{2}\int_{-1}^1d{\cal O}\, \Sigma^2_\text{mix}({\cal O}') = \frac{1}{2}+\frac{2\epsilon}{\pi^2}\,.
\end{align}
Clearly, when $\epsilon = 0$, there is no discrimination power, and when there is maximal discrimination power, $\epsilon = 1$ and the AUC is
\begin{align}
\text{AUC}(\epsilon=1) =  \frac{1}{2}+\frac{2}{\pi^2} \approx 0.7026\,,
\end{align}
satisfying our general bound.  The value of the KS test is
\begin{align}
\text{KS} =  \int_{-1}^0 d{\cal O}\, 
\Sigma_\text{mix}({\cal O}) = \frac{\epsilon}{\pi}\,,
\end{align}
again satisfying our general bound.  The KL divergence can be calculated analytically as a function of $\epsilon$, but its form is not illuminating.  However, its maximum value, when $\epsilon = 1$ is
\begin{align}
D_\text{KL}(p_\text{pure}||p_\text{mix}) \leq 1-\log 2 \,,
\end{align}
satisfying the general bound.

\subsection{Gluon Helicity Interference}

For our first physical example, we consider the interference of helicity states of an intermediate gluon studied in \Ref{Chen:2020adz}.  In general, an intermediate gluon in a Feynman diagram calculation is not physical because its virtuality is non-zero and its representation depends on the choice of gauge.  However, in the limit in which an intermediate gluon splits to two massless, collinear particles, the intermediate gluon becomes on-shell and real.  As it is real, it has two physical helicity states, but because it has split to two other particles, the intermediate gluon is not directly observed.  Therefore, in a calculation, we must sum coherently over the two orthogonal helicity states of the intermediate gluon, and when the amplitude is squared, this results in interference that is imprinted on the detected daughter particles.

The procedure presented by \Ref{Chen:2020adz} for observing this interference is the following.  One considers the measurement of the three-particle energy correlator \cite{Chen:2019bpb}, which sums over all triples of particles, weighted by the product of the three energies in the triple, for fixed pairwise relative angles between particles.  To ensure that the particle that is probed is (nearly) on-shell, the angle between two of the particles in the triple is taken very small with respect to the other two angles, called the squeezed limit.  Then, the azimuthal angle $\phi$ that the squeezed particles make about their center-of-mass with respect to the third, wide-angle particle can be measured.  The interference of the two intermediate gluon spin states is imprinted on the azimuthal distribution of its squeezed daughter particles.  

We refer to \Refs{Chen:2020adz,Chen:2019bpb} for all details and complete calculations, but here we will just extract one azimuthal distribution established there.  In a collider experiment, we never directly observe the daughter quarks or gluons, so we must sum together all possible final states.  Unlike for the spin states of the intermediate gluon, the sum over final states is incoherent, because we could in principle make a measurement of the final state particle flavor.  For an intermediate gluon emitted off of a hard quark, the normalized, leading-order, squeezed azimuthal distribution that exhibits quantum interference can be expressed as
\begin{align}
p_\text{pure}(\phi)&=\frac{2}{\pi}+\frac{20(C_A-2n_f T_R)}{3\pi(91 C_A+240 C_F+13 n_f T_R)}\cos(2\phi)\,.
\end{align}
This is normalized on $\phi\in[0,\pi/2]$, and the interference manifests as $\cos(2\phi)$ because the difference between the helicity $+1$ and helicity $-1$ intermediate gluons is spin-2.  In this expression, $C_F=4/3$ and $C_A=3$ are the fundamental and adjoint quadratic Casimirs of SU(3) color, respectively, $T_R = 1/2$, and $n_f$ is the number of quarks to which the intermediate gluon could split.  With $n_f = 5$, the approximate numerical value of the pure state distribution is
\begin{align}
p_\text{pure}(\phi)&\approx \frac{2}{\pi}-0.006785\cos(2\phi)\,,
\end{align}
and so the interference has an extremely small amplitude.  This large suppression arises because of an unfortunate cancelation between the interference of quark and gluon products from the splitting as a consequence of approximate supersymmetry, and could perhaps be increased if the flavor purity of the final state could be improved.  Nevertheless, the interference is non-zero and so the pure state distribution is different from the mixed state distribution
\begin{align}
p_\text{mix}(\phi)&=\frac{2}{\pi}\,.
\end{align}

The optimal discrimination observable is then
\begin{align}
{\cal O} &=\frac{10(C_A-2n_f T_R)}{3(91 C_A+240 C_F+13 n_f T_R)}\cos(2\phi)\\
&\approx -0.01066\cos(2\phi)\,.\nonumber
\end{align} 
Its distribution on the mixed and pure samples take the same form as for any sinusoidal interference, where here
\begin{align}
p_\text{mix}({\cal O}) \approx \frac{1}{\pi\sqrt{(0.01066)^2-{\cal O}^2}}\,,
\end{align}
where we identify $\epsilon \approx 0.01066$.  Then, the AUC is
\begin{align}
\text{AUC} \approx 0.5022\,,
\end{align}
very slightly over completely random.  The KS test value is
\begin{align}
\text{KS} \approx 0.003393\,.
\end{align}
The KL divergence, or the information needed to encode the interference, is also tiny,
\begin{align}
D_\text{KL}(p_\text{pure}||p_\text{mix}) \approx 2.841 \times 10^{-5}\,.
\end{align}

\subsection{Spin-0 and Spin-2 Higgs Interference}

For our final example of quantum interference, we will consider the possibility that there exists both a spin-0 Higgs boson as well as a spin-2 Higgs boson.  If there are multiple Higgs bosons in nature of different spin, then they can both in principle contribute to processes like $gg\to \gamma\gamma$ that are fundamental for establishing the existence of the Higgs.  For some helicity configurations of the initial gluons and final photons, there is non-zero probability for both the spin-0 and spin-2 Higgs to mediate the process, and so they must be coherently summed in an amplitude for a complete prediction.  Higgs bosons of different spin are orthogonal states, and so when the amplitude is squared, this results in quantum interference that can in principle be observed.

To isolate and study this interference of different spin Higgs bosons, we have to assume that the helicities of the gluons and photons can be determined.  This is of course unrealistic because at a hadron collider like the LHC, gluons are extracted from protons and their spin is mixed with all other constituents of the proton.  Additionally, the spin of a final state photon is not observable at the LHC because there are no polarizing filters in the experiments, for example.  While unrealistic, this will illustrate other interesting features of pure versus mixed state distributions that were not present in the previous example.

For concreteness, we will consider the scattering process $g^+g^+\to \gamma^+\gamma^+$, where all external particles are massless have +-helicity.  In the center-of-mass frame, the gluons collide head-on and the photons travel out of the collision back-to-back.  The only kinematic quantity that the distribution on phase space can depend on is the scattering angle $\theta$, the angle between one of the initial gluons and one of the final photons.  Further, the distribution of the scattering angle $\theta$ must be a Legendre polynomial in $\cos\theta$, which are the partial waves or representations of SO(3) rotations (with no azimuthal dependence).  Therefore, the scattering amplitudes for this process, mediated by either a spin-0 or spin-2 Higgs are:
\begin{align}
{\cal A}(g^+g^+\to H_\text{spin-0}\to \gamma^+\gamma^+)& \propto P_0(\cos\theta) = 1\,,\\
{\cal A}(g^+g^+\to H_\text{spin-2}\to \gamma^+\gamma^+)&\propto  P_2(\cos\theta) \propto1-3\cos^2\theta\,,\nonumber
\end{align}
where $P_n(x)$ is the $n$th Legendre polynomial.  The normalized pure and mixed state distributions on phase space are then
\begin{align}
d\cos\theta\, \langle \rho_\text{pure}\rangle &= \frac{1}{2}\frac{1}{1+\epsilon}+\frac{5}{8}\frac{\epsilon}{1+\epsilon}(1-3\cos^2\theta)^2\\
&
\hspace{2cm}+\sqrt{\frac{10}{8}}\frac{\sqrt{\epsilon}}{1+\epsilon}(1-3\cos^2\theta)\,,\nonumber\\
d\cos\theta\,\langle \rho_\text{mix}\rangle &= \frac{1}{2}\frac{1}{1+\epsilon}+\frac{5}{8}\frac{\epsilon}{1+\epsilon}(1-3\cos^2\theta)^2\,.\nonumber
\end{align}
In these expressions, we use the shorthand notation $\langle \rho\rangle$ to denote the density matrix projected onto the appropriate $N$-particle momentum state.  As always, $\epsilon$ quantifies the relative probability of the contribution of the spin-0 or spin-2 Higgs to the process.

From these distributions, the discrimination observable is the ratio of the interference term to the mixed state distribution:
\begin{align}
{\cal O} = \frac{\sqrt{80}\sqrt{\epsilon}(1-3\cos^2\theta)}{4+5\epsilon(1-3\cos^2\theta)^2}\,.
\end{align}
Its distribution on the mixed state can be calculated in the usual way:
\begin{align}
p_\text{mix}({\cal O}) &= \int_0^1 \frac{dx}{\sqrt{x}}\left(
\frac{1}{2}\frac{1}{1+\epsilon}+\frac{5}{8}\frac{\epsilon}{1+\epsilon}(1-3x)^2
\right)\\
&
\hspace{2cm}\times\delta\left(
{\cal O}-\frac{\sqrt{80}\sqrt{\epsilon}(1-3x)}{4+5\epsilon(1-3x)^2}
\right)\,,\nonumber
\end{align}
where we have made a change of variables $x = \cos^2\theta$ to render the expression more compact.  This integral can be explicitly evaluated and one finds 
\begin{align}
&p_\text{mix}({\cal O}) =\frac{2}{\sqrt{\epsilon}(1+\epsilon)}\,\Theta\left(
\frac{4\sqrt{5\epsilon}}{4+5\epsilon}-{\cal O}
\right)\\
&\hspace{1cm}\times\frac{(1-\sqrt{1-{\cal O}^2})^3}{{\cal O}^4\left(
{\cal O}^2-1+\sqrt{1-{\cal O}^2}
\right)\sqrt{15-6\sqrt{5}\frac{1-\sqrt{1-{\cal O}^2}}{\sqrt{\epsilon}{\cal O}}}}\nonumber\\
&
\hspace{1cm}
+\frac{2}{\sqrt{\epsilon}(1+\epsilon)}\,\Theta\left(
-\frac{2\sqrt{5\epsilon}}{1+5\epsilon}-{\cal O}
\right)\nonumber\\
&
\hspace{1cm}
\times\frac{(1+\sqrt{1-{\cal O}^2})^3}{{\cal O}^4\left(
1-{\cal O}^2+\sqrt{1-{\cal O}^2}
\right)\sqrt{15-6\sqrt{5}\frac{1+\sqrt{1-{\cal O}^2}}{\sqrt{\epsilon}{\cal O}}}}
\nonumber\,.
\end{align}
While a bit unwieldy, this distribution lacks the ${\cal O}\to -{\cal O}$ symmetry that we had observed in the case of gluon helicity interference.  For illustration, we will just present the value of the KL divergence in the case of maximal mixing $\epsilon = 1$ of the different Higgs bosons. We find
\begin{align}
D_\text{KL}(p_\text{pure}||p_\text{mix}) \approx 0.473\,,
\end{align}
which is indeed less than the upper bound $\log 2 \approx 0.6931$.

\section{Conclusions}\label{sec:conc}

Observing signatures of quantum mechanics at particle physics colliders is challenging because of the necessary way that data are collected.  We established general results for observation of quantum interference at a collider arising from coherent summation of orthogonal states to the amplitude of a process.  Measurements at a collider take place on the space of $N$-particle momentum states, so direct measurement of off-diagonal elements of the density matrix of a system are not possible.  Instead, pure and mixed states have distinct probability distributions on phase space, and their likelihood ratio defines the optimal observable for identification of interference.  These results enable general, robust bounds that set limits on the feasibility of observing interference.

For simplicity and clarity, our analysis here has been limited in a few ways.  First, we only considered the interference of two orthogonal states on phase space.  Second, in the explicit examples we considered to illustrate quantum interference, we restricted to cases in which the interfering states projected onto only a single particle number state.  It would be interesting to generalize these results.  When multiple orthogonal states interfere, the interference still only arises through relative phases of pairs of states, but now there would be a sum over many such pairs.  Our results likely naturally generalize for this multiple state interference case.  On the other hand, analyzing states that interfere that project onto multiple particle number states, or even an arbitrary numbers of them, may be significantly different and challenging.  Numerous systems that appear in particle physics consist of arbitrary numbers of particles, most notably arising due to the presence of infrared singularities or approximate scale invariance.  Jets, collimated streams of hadrons arising from dynamics of the strong force at high energies, are perhaps the most prominent objects that lack a well-defined particle number, and determination of the manifestation of interference on jets could be a powerful probe of quantum mechanics in a vastly different regime.  Recently, methods for defining entropy on jets has been performed \cite{Neill:2018uqw}, but the authors note that the dynamics of a jet itself lead to decoherence and may wash out nearly all observable traces of interference.

As mentioned in the introduction, establishing Bell-like inequalities for observation of entanglement is extremely subtle at a collider.  However, it is interesting to ask if similar inequalities can be established for quantum interference.  If the density matrix could be measured at a collider, this would be straightforward, if technically challenging, because the von Neumann entropy could be determined and the extent to which an event is a mixed state quantified.  Without direct access to the density matrix, and only access to squared scattering amplitudes on fixed $N$-body phase space, a path forward to such a goal is much murkier.  Nevertheless, it could shine a light on the strange ways that the universe works at the shortest distances probed by the highest energy colliders.

\begin{acknowledgements}
This work was supported by the Department of Energy, Contract DE-AC02-76SF00515.

\end{acknowledgements}

\bibliography{quantint}

\end{document}